\documentclass[3p,authoryear,preprint,12pt]{elsarticle}
\makeatletter\if@twocolumn\PassOptionsToPackage{switch}{lineno}\else\fi\makeatother

\usepackage{tabulary,xcolor}
\usepackage{amsfonts,amsmath,amssymb}
\usepackage[T1]{fontenc}

\usepackage{enumitem}
\usepackage{amssymb} 
\usepackage{graphicx}
\makeatletter
\let\save@ps@pprintTitle\ps@pprintTitle
\def\ps@pprintTitle{\save@ps@pprintTitle\gdef\@oddfoot{\footnotesize\itshape \null\hfill\today}}
\def\hlinewd#1{%
  \noalign{\ifnum0=`}\fi\hrule \@height #1%
  \futurelet\reserved@a\@xhline}
\def\tbltoprule{\hlinewd{.8pt}\\[-12pt]}
\def\tblbottomrule{\noalign{\vspace*{6pt}}\hline\noalign{\vspace*{2pt}}}

\AtBeginDocument{\ifNAT@numbers \biboptions{sort&compress}\fi}

\makeatother

\usepackage[utf8]{inputenc} 
\ifluatex

\usepackage[T1]{fontenc} 
\defaultfontfeatures{Ligatures=TeX}
\usepackage{lmodern}
\unimathsetup{math-style=TeX}
\else 
\usepackage[utf8]{inputenc}
\fi 
\ifluatex\else\usepackage{stmaryrd}\fi

\usepackage{url,multirow,morefloats,floatflt,cancel}
\makeatletter

\AtBeginDocument{\@ifpackageloaded{textcomp}{}{\usepackage{textcomp}}}
\makeatother
\usepackage{colortbl}
\usepackage{xcolor}
\usepackage{pifont}
\usepackage{pgf-pie}
\usepackage[nointegrals]{wasysym}
\makeatletter

\def\mcWidth#1{\csname TY@F#1\endcsname+\tabcolsep}

\def\cAlignHack{\rightskip\@flushglue\leftskip\@flushglue\parindent\z@\parfillskip\z@skip}
\def\rAlignHack{\rightskip\z@skip\leftskip\@flushglue \parindent\z@\parfillskip\z@skip}

\@ifundefined{etal}{}{}

\usepackage{ifxetex}
\ifxetex\else\if@twocolumn\@ifpackageloaded{stfloats}{}{\usepackage{dblfloatfix}}\fi\fi

\AtBeginDocument{
\expandafter\ifx\csname eqalign\endcsname\relax
\def\eqalign#1{\null\vcenter{\def\\{\cr}\openup\jot\m@th
  \ialign{\strut$\displaystyle{##}$\hfil&$\displaystyle{{}##}$\hfil
      \crcr#1\crcr}}\,}
\fi
}

\AtBeginDocument{%
  \@ifpackageloaded{endfloat}%
   {\renewcommand\efloat@iwrite[1]{\immediate\expandafter\protected@write\csname efloat@post#1\endcsname{}}}{\newif\ifefloat@tables}%
}%

\def\BreakURLText#1{\@tfor\brk@tempa:=#1\do{\brk@tempa\hskip0pt}}
\let\lt=<
\let\gt=>
\def\processVert{\ifmmode|\else\textbar\fi}

\@ifundefined{subparagraph}{
\def\subparagraph{\@startsection{paragraph}{5}{2\parindent}{0ex plus 0.1ex minus 0.1ex}%
{0ex}{\normalfont\small\itshape}}%
}{}

\newcommand\role[1]{\unskip}
\newcommand\aucollab[1]{\unskip}
  
\@ifundefined{tsGraphicsScaleX}{\gdef\tsGraphicsScaleX{1}}{}
\@ifundefined{tsGraphicsScaleY}{\gdef\tsGraphicsScaleY{.9}}{}
\def\checkGraphicsWidth{\ifdim\Gin@nat@width>\linewidth
	\tsGraphicsScaleX\linewidth\else\Gin@nat@width\fi}

\def\checkGraphicsHeight{\ifdim\Gin@nat@height>.9\textheight
	\tsGraphicsScaleY\textheight\else\Gin@nat@height\fi}

\def\fixFloatSize#1{}
\let\ts@includegraphics\includegraphics

\def\inlinegraphic[#1]#2{{\edef\@tempa{#1}\edef\baseline@shift{\ifx\@tempa\@empty0\else#1\fi}\edef\tempZ{\the\numexpr(\numexpr(\baseline@shift*\f@size/100))}\protect\raisebox{\tempZ pt}{\ts@includegraphics{#2}}}}

\AtBeginDocument{\def\includegraphics{\@ifnextchar[{\ts@includegraphics}{\ts@includegraphics[width=\checkGraphicsWidth,height=\checkGraphicsHeight,keepaspectratio]}}}

\DeclareMathAlphabet{\mathpzc}{OT1}{pzc}{m}{it}

\def\URL#1#2{\@ifundefined{href}{#2}{\href{#1}{#2}}}

\def\UrlOrds{\do\*\do\-\do\~\do\'\do\"\do\-}%
\g@addto@macro{\UrlBreaks}{\UrlOrds}

\edef\fntEncoding{\f@encoding}

\makeatother

\newif\ifmultipleabstract\multipleabstractfalse%
    \makeatletter
\def\ead{\@ifnextchar[{\@uad}{\@ead}}
\gdef\@ead#1{\bgroup
   \def\_{\string\underscorechar\space}
   \def\{{\string\lbracechar\space}
   \def\textdagger{\string\textdagger\space}
   \def\texttildeapprox{\string\texttildeapprox\space}
   \def~{\hashchar\space}
   \def\}{\string\rbracechar\space}
   \edef\tmp{\the\@eadauthor}
   \immediate\write\@auxout{\string\emailauthor
     {#1}{\expandafter\strip@prefix\meaning\tmp}}
  \egroup
}
\gdef\emailauthor#1#2{\stepcounter{ead}
      \g@addto@macro\@elseads{\raggedright
      \let\corref\@gobble
      \eadsep\texttt{#1} (#2)
      \def\eadsep{\unskip,\space}}
}

\makeatother
  
\usepackage{float}

\begin{document}

\nocite{*}

\begin{frontmatter}

    \title{
  Investigating the Impact of Project Risks on Employee Turnover Intentions in the IT Industry of Pakistan    
}
    
\author[a8e6b2fa14c62]{Ghalib Ahmed Tahir} \author[a0945b41fa13d]{Murtaza Ashraf}

\begin{abstract}
Employee turnover remains an ongoing issue within high-tech sectors such as IT firms and research centers, where organizational success relies on the skills of their workforce. Intense competition and a scarcity of skilled professionals in the industry contribute to an ongoing demand for highly qualified employees, posing challenges for organizations to retain talent. While numerous studies have explored various factors affecting employee turnover in these industries, their focus often remains on overarching trends rather than specific organizational contexts. In particular, within the software industry, where project-specific risks can affect project success and timely delivery, understanding their influence on job satisfaction and turnover intentions is relevant. This study aims to examine the influence of project risks in the IT industry on job satisfaction and employee turnover intentions. Furthermore, it examines the role of both external and internal social links in shaping perceptions of job alternatives (PJA).
\end{abstract}
    
  \end{frontmatter}

\section{Introduction}
Employee turnover presents a notable challenge in the high-tech industry, where skilled personnel are important for success. This is particularly evident in the IT sector, where employees play an important role throughout various project phases, from requirement gathering to development. The departure of main employees can disrupt projects, leading to delays or even failure. While early perspectives attributed turnover to job dissatisfaction, contemporary research has expanded its scope to include factors like organizational commitment, climate, and job alternatives. \mbox{}\protect\newline Aggregate studies suggest that labor market conditions influence overall turnover rates \unskip~\cite{75323:2894213}. However, when scrutinizing individual decisions to stay or leave an organization, researchers find that general unemployment rates do not exert a direct impact\unskip~\cite{75323:2894214}. Moreover, turnover rates vary across different departments, with industries like accommodation and employee services experiencing higher turnover (around 50\%) compared to sectors like software (around 20\%) or educational services (over 10\%). \mbox{}\protect\newline Not all turnovers carry the same weight. Initially, involuntary turnover was believed to be within the organization's control, while uncontrolled turnover presented unique challenges that couldn't be fully integrated into turnover models. However, managers recognize that not all turnover is detrimental; the departure of underperforming employees may benefit organizational results, whereas losing high-performing individuals can significantly impact productivity. Hence, it becomes relevant for organizational management to discern between preventable and unpreventable turnover. This distinction allows them to devise targeted strategies to retain valuable talent and mitigate the adverse effects of turnover on organizational outcomes.\unskip~\cite{75323:2894216}.

From an economic standpoint, the costs associated with employee turnover are important for organizations. Presently, financial statements do not explicitly account for the costs incurred due to voluntary turnover. Instead, these costs are distributed across various processes like recruitment, selection, training, and mentoring. Moreover, the impact of turnover on organizational knowledge continuity is often overlooked. The estimated loss per departing employee can range from thousands of dollars to more than double the employee's salary, taking into account the loss of organizational knowledge. \mbox{}\protect\newline Turnover rates also vary across different countries globally. For instance, there are perceptible differences between turnover rates in the United States and European countries. Research findings suggest that Europeans are approximately half as likely to change jobs within a year compared to Americans. In summary, voluntary employee turnover stems from various causes, and maintaining a high employee retention rate is relevant for maintaining a competitive edge over other organizations. \mbox{}\protect\newline Predicting turnover based on macroeconomic data is feasible due to decades of research, yet understanding individual behavior amidst constantly evolving conditions requires ongoing research efforts. There is a need to investigate factors specific to particular industries in developing countries, as much of the existing research is based on Western data samples. Additionally, exploring the applicability of these findings in diverse cultural and societal contexts is essential. \mbox{}\protect\newline In our research, we have examined various relatively unexplored factors and their relationships that influence employee turnover, particularly within the IT industry. Given the notable risks associated with IT projects, such as strict deadlines and unclear requirements, project risk factors play a important role in project execution and delivery. These factors can also contribute to job dissatisfaction and subsequent employee turnover. \mbox{}\protect\newline Furthermore, we have studied the impact of both external/internal social networks on perceived job alternatives (PJA), as well as the moderating role of perceived job alternatives between job satisfaction and turnover. We proposed a structured questionnaire to assess the influence of these factors within the dynamic IT industry, which was reviewed by project managers and researchers. In our results, we have discussed the influence of these factors mentioned above on employee turnover intentions.
    
\section{Research Objective}
The following are the aims of this research: \mbox{}\protect\newline 

\begin{itemize}
  \item \relax To examine the impact of project-specific risks on employee turnover in the context of IT corporations.
  \item \relax To examine the mediating role of internal or social links through perceived job alternatives and moderating the relationship in satisfaction of employee job and turnover intentions.
\end{itemize}

\section{Literature Review}
In this section, we define main concepts drawn from existing literature and outline the theories pertaining to employee turnover intentions. Subsequently, we have presented our main ideas that form the basis of the theoretical framework.

\subsection{\textbf{Theories of Employee Turnover Intentions}}In the past Employee, turnover has been extensively studied. Various theories are proposed in the past. These theories are listed below.

a) Theory of Organization Equilibrium\unskip~\cite{75323:2894201}

b) Met Expectation Model \unskip~\cite{75323:2894202}

c) Turnover Process Model \unskip~\cite{75323:2894203}

d) Revised Intermediate {\textemdash} Process Model \unskip~\cite{75323:4296950}

e) Structural Model \unskip~\cite{75323:4296951}

f) Expanded Model \unskip~\cite{75323:2894205}

g) Multidisciplinary Model \unskip~\cite{75323:2894265}

h) Ferrell and Rusbult: Investment Model \unskip~\cite{75323:2894266}

i) Steers and Mowday: Multi {\textemdash} Route Model 

J) Cusp Catastrophe Model \unskip~\cite{75323:2894267}

k) Labor Economic Model \unskip~\cite{75323:2894268}

l) Unfolding Model\unskip~\cite{75323:4302196}

m) Job embeddedness theory \unskip~\cite{75323:2894236}

The trends of employee turnover in the past decade are more focused on the following factors 

a) Turnover predictions of new individual differences.  

b) Increase significance on contextual variables and giving importance t to interpersonal relationship 

c) Increased focus on aspects at residing especially organizational responsibility and job-embeddedness 

d) vigorous modeling of turnover system with consideration of time. 

The current trends of the studies in the area of employee turnover intentions includes the following

a) Influence of social networks including community embeddedness and organizational embeddedness 

b) Temporary (early vs. late turnover) 

c) Turnover Consequences

d) Multilevel turnover investigations

e) Other withdrawal types e.g. retirement.

In the current competitive environment, attracting and retaining top talent presents a notable challenge for managers. With the expansion of knowledge work, globalization, and technological advancements, the importance of employee retention has heightened compared to previous eras. To mitigate intentional turnover and overall turnover rates, managers implement human resource policies aimed at retaining valuable employees  \unskip~\cite{75323:2894209,75323:2894210}\unskip~\cite{75323:2894211}. Despite the substantial body of research, with over 1500 papers addressing the subject, turnover remains a dynamically evolving field of study. New administrative strategies, changes in job market dynamics, advancements in research methodologies, and technology have all contributed to the evolving nature of turnover research.

a) Time period before 1985

b) Time period from 1985 to 1995

c) Advances from 1996 to present

d) Present advancement of employee turnover in context with information management

\subsection{\textbf{Time Period Before 1985}}There are several notable models, proposed before 1985. These models laid the foundation for future studies in employee turnover.

  \begin{enumerate}
  \item \relax \unskip~\cite{75323:2894201}
  \item \relax \unskip~\cite{75323:2894202}
  \item \relax \unskip~\cite{75323:2894203}
  \item \relax \unskip~\cite{75323:2894205}
  \item \relax \unskip~\cite{75323:4301328}
  \item \relax \unskip~\cite{75323:2894276}
  \item \relax \unskip~\cite{75323:4296950}
  \end{enumerate}
  One of the influential theoretical contributions in this field came from \unskip~\cite{75323:2894201}. Their relevant work introduced the theory of "organizational equilibrium" in their book published in 1958. They have emphasized the importance of balancing employees, contributions of the organization, and inducements. Central to their theory are two main factors influencing employee harmony: perceived attractiveness and perceived ease of leaving the organization. In contemporary literature, these concepts are commonly referred to as job satisfaction and perceived job alternatives.

\subsection{\textbf{Turnover Research From 1985 to 1995}}From 1985 to 1995, research in the field primarily shifted its focus towards complex organizational and group-level concepts, such as organizational environment, group cohesion, gender composition, and demographics.Unlike the preceding era, which mainly concentrated on individual-level factors, attention expanded to include contextual variables that could have a observable impact on employee turnover. These contextual variables were further categorized into two main groups:

\begin{itemize}
  \item \relax Organization macro-level variables.
  \item \relax Person Context incorporates values with attention to employee's relationship with the climate.
\end{itemize}
   \mbox{}\protect\newline According to \unskip~\cite{75323:2894217}, organizational culture plays a important role in turnover by shaping employees' perceptions, leading to withdrawal behaviors. Another study by \unskip~\cite{75323:2894218} found that pay inequality among university administrators predicts employee turnover, while differences in tenure at the company decrease social integration and influence individual turnover. In addition to macro-level variables like organizational culture and pay systems, researchers also examined other factors such as organizational fit, network centrality, and mentoring. \unskip~\cite{75323:2894219} concluded that low person-organization fit after 20 months leads to increased turnover. From a social network perspective, \unskip~\cite{75323:2894275} examined that individuals with more connections within a department are less likely to turnover compared to others. \mbox{}\protect\newline Another notable advancement during this period was the three-angle conceptualization of organizational commitment by \unskip~\cite{75323:2894220}. Empirical testing of the three facets of this conceptualization further assessed their contributions to understanding turnover-related withdrawal attitudes and behaviors. Researchers also began incorporating recent behaviors into turnover studies, particularly those related to stress, well-being, and ambiguity. Job insecurity and emotional exhaustion were found to be positively linked to turnover intentions. \mbox{}\protect\newline Towards the end of the 1985-1995 period, Lee and Mitchell introduced a new direction to researchers by unveiling the model of the turnover process. Drawing from image theory, they proposed that turnover intentions are not solely driven by job dissatisfaction but may also occur without much consideration. They outlined five pathways that employees may pursue before making a turnover decision, with Path 1 being distinct from previous models as it involves quitting the organization due to environmental shock rather than dissatisfaction. This shock, defined as a jarring event, triggers the psychological analysis leading to job quitting. \mbox{}\protect\newline There are two other paths which are initiated by shocks while the rest two are traditional paths driven by job dissatisfaction. \mbox{}\protect\newline Overall, this model underscores the dynamics and complexity of the turnover process, suggesting that future research should consider how people leave their jobs. An interesting research contribution is the integrative adaptation and withdrawal model proposed by \unskip~\cite{75323:2894207}. This model suggests that job dissatisfaction, whether general or specific, triggers a sequence of cognitive and behavioral responses leading to adaptive behaviors. The model argues that withdrawal attitudes are interconnected, and withdrawal behaviors should be viewed as part of a broader pattern encompassing various attitudes, contrasting with previous studies that treated behaviors like turnover and lateness independently. \mbox{}\protect\newline In conclusion, researchers have broadened their perspective by examining individual-level consequences such as strain and organizational-level consequences like increased turnover among remaining employees at the next job \unskip~\cite{75323:2894250}.

\subsection{\textbf{Turnover Research From 1995 To Present}}In the previous decade, there is a considerable theoretical expansion in the area of employee turnover. There are seven notable trends of employee turnover in the last decade. All of these trends are described below.

\begin{itemize}
  \item \relax Introduction of new change forecasts for turnover.
  \item \relax Increased emphasis on stress and change-related perspectives.
  \item \relax Empirical analysis of unfolding models.
  \item \relax Heightened attention to contextual parameters, particularly social relationships.
  \item \relax Enhanced focus on factors influencing employee retention.
  \item \relax Dynamic modeling of the turnover process with consideration of time.
  \item \relax Expansion of previously identified relationships.
\end{itemize}
  Various theoretical frameworks have contributed to our understanding of employee turnover, enriching the field. However, there remains a gap in comprehending the cooperative aspect of the turnover process. Some studies have delved into individual difference predictors of turnover, exploring both direct and moderating effects. For instance, some researchers have found that personality traits may influence an individual's decision to leave their job. According to \unskip~\cite{75323:2894216}, a person's confidence and decision-making, when combined with their bio-data, are negatively associated with turnover during the recruitment phase. Additionally, \unskip~\cite{75323:2894215} showed that one of the notable personality traits, scrupulousness, is negatively correlated with turnover.

Research articles published by \unskip~\cite{75323:4301510} suggest that negative affectivity increases both the intention of leaving and actual turnover. Similarly, \unskip~\cite{75323:2894221} found that employees with low self-monitoring tendencies and risk reluctance are more likely to turn their desire to leave into actual turnover.

In essential studies aiming to merge different abstract approaches, \unskip~\cite{75323:4301594} has connected agreeableness and proposed various models. They have showed that their eight turnover factors, previously presented, are systematically associated with four decision categories of turnover. Different groups of job quitters are usually prompted by different forces.

They have identified eight different factors of turnover cognition, which serve as the leading predictors of turnover management. Furthermore, they proposed that these factors intervene in the effects of all other main constructs in the literature. One notable finding of their empirical testing is that individuals who quit their jobs without other alternatives have a more negative impact compared to users of other decision types, implying effect-driven, impulsive quitting.

\unskip~\cite{75323:4301595} published their results on the moderating aspect of extraversion on leader-member interchange and turnover association during new managerial advancement. They found that individuals with low extraversion exhibit a notable negative association between leader-member exchange (LMX) and executive leaving compared to those with high extraversion.

Maertz and their team introduced a single framework that aligns many individual designs to impact turnover, which is quite useful in the research area. With increased importance on organizational change and employee adaptation, rapidly changing environments and stress attitudes have become notable areas of behavioral research concerning employee turnover. For example, psychological uncertainty, explored as a turnover predictor, was positively linked with turnover intentions and influenced by changing frequency and planning involved in change, as well as transformational change\unskip~\cite{75323:2894222}. Change acceptance was positively linked with job satisfaction and negatively linked with work displeasure and turnover intentions, aiding in predicting real turnover. Additionally, transformational change positively relates to employee turnover by emphasizing the value retention of employees managed during times of rapid organizational change.

Despite previous researchers including stressors in their models, current research has considered different types of stressors and investigated their benefits. Consistent with previous era of research, hindrance stressors such as role conflict, organizational politics, role overload, and situational constraints are negatively associated with job satisfaction and organizational commitment, while positively related to employee turnover and turnover intentions. On the other hand, challenge stressors like time urgency and task pressure are positively linked to job attitudes and negatively related to turnover intentions \unskip~\cite{75323:2894223}\unskip~\cite{75323:2894257}.

Researchers have also explored stressful events. For instance, Iverson and Pullman found that voluntary turnover and layoffs have different antecedents. Factors which influenced the willingness to leave a job include blue-collar status versus white-collar status (negative correlation), age (negative correlation), intent to leave (positive correlation), and shock (positive correlation). In contrast, layoffs are forecasted by factors such as full-time status versus part-time status (positive correlation), age (positive correlation), and absenteeism (negative correlation). Additionally, \unskip~\cite{75323:2894224} stated that there is a positive relation between voluntary turnover and supervisory abuse, mediated by perceptions of organizational justice and moderated by perceived mobility. This relationship is stronger for employees with lower perceived mobility. Furthermore, research by\unskip~\cite{75323:2894225} has shown that harassment has a negative impact on turnover.

\unskip~\cite{75323:2894208} have argued for the need of an alternative theory to explain why and how people leave their companies.They proposed the unfolding model, whose main constituents are shock, script, violations of image, job search, and job satisfaction.Firstly, shock refers to an event that triggers the psychological analysis involved in quitting a job.Aftershock script is a preexisting plan of action required for leaving the job.Image violations occur when personal goals do not align with the organization's goals and strategies.Additionally, job satisfaction decreases when individuals feel their job does not provide desired technical skills, financial benefits, or knowledge.Finally, job search involves seeking alternative job opportunities and evaluating them.These components of the model evolve over time and integrate to form five unique pathways for individuals to leave their jobs.

In Path 1, shock initiates the execution of a preexisting plan where the individual does not consider alternative job opportunities or their association with the organization. In Path 2, shock prompts the individual to reconsider their association with the organization due to image violations, leading them to leave without considering alternative job opportunities. Path 3 involves shock violating the image of employees, prompting them to compare their current job with other job offers. In Path 4, lower job satisfaction causes the individual to leave the organization without considering alternative job opportunities. Many studies have shown that shock is a notable factor leading individuals to leave their organization. According to \unskip~\cite{75323:2894277}, shock is the most immediate cause of employee turnover compared to job satisfaction. 

Furthermore, a replication of experiments conducted by \unskip~\cite{75323:2894279} extended previous tests on the unfolding model, successfully classifying 86\% of samples who left their organization. Their results suggest that most people leave or stay due to economic considerations. Additionally, they found that women experience shock more than men and tend to follow Paths 1, 2, or 3. Research conducted by \unskip~\cite{75323:2894226} in Great Britain found that 44\% of nurses reported that shocks impacted their decision to turnover.

In summary, there is empirical evidence supporting the main concepts of the unfolding model, particularly the concept of shock. However, most of these studies have been conducted on Western data, highlighting the need for similar studies in Eastern contexts.

There are also many research studies which take into account contextual considerations.

\subsection{\textbf{Organizational Context/ Macro Level}}Researchers have introduced additional organizational context variables such as unit-level attitudes, engagement, satisfaction, and climate approach. Employee engagement and satisfaction across various business units are not positively associated with employee turnover\unskip~\cite{75323:2894227}. Furthermore, \unskip~\cite{75323:2894258} have reported that the performance of a unit is negatively affected by both involuntary and voluntary turnover. When controlling for the other two schemes, each scheme has negative consequences, albeit with different effects. \unskip~\cite{75323:2894228} has stated that citizenship and level of unit satisfaction behavior from the first year positively forecast satisfaction of customers and unit profit in the second year. On the other hand, turnover from the first year outcome has negatively predicted second-year results.\unskip~\cite{75323:2894211} has shown that turnover at the unit level is negatively linked with unit performance. This connection was mediated by unit efficiency, indicating that workforce stability is required for organizational outcomes.

Researchers have also examined the impact of organizational diversity on employee turnover. Elvira and Cohen investigated the effect of gender composition on one's own gender at various positions within the organization. Their results indicated that employees of the same sex had an impact on the retention of female employees, negatively affecting the turnover rate, but had no impact on male employees. Research on public accountants conducted by \unskip~\cite{75323:2894274} reported that female public accountants are more likely to leave their positions compared to their male counterparts, not due to family issues but because of the lack of promotional opportunities. Conversely, research results by\unskip~\cite{75323:2894229} showed that female financial managers were less likely to leave compared to male financial managers when they obtained a higher position in the preceding eleven months. Demographic misfit within work groups is also associated with a high turnover hazard. A comprehensive study conducted by\unskip~\cite{75323:2894237} on more than 20 US corporations and 450,000 professionals and managers found that incumbents of jobs typically held by African Americans and Hispanics were at a higher risk of turnover. Other organizational variables, such as firms' pay distribution, also affect employee turnover\unskip~\cite{75323:2894253}. \unskip~\cite{75323:2894259} found that the negative effect of perceived supervisory support on employee turnover is mediated by perceived organizational support. However, a study by \unskip~\cite{75323:2894231} showed that perceived supervisor support has an independent effect on turnover cognition, and perceived organizational support had a notable impact on turnover mediated through affective and organizational commitment.

\subsection{\textbf{Person Context Interface}}Current research in this perspective focuses extensively on interpersonal relations and the interface between employees and their environment, a dimension largely overlooked by previous models from the 1990s era. In recent studies, turnover has been examined from a relational perspective. For instance,\unskip~\cite{75323:2894251} studied the impact of minority network associations on minority employee turnover, highlighting the significance of social embeddedness in understanding turnover dynamics. Similarly, research by \unskip~\cite{75323:2894260}\unskip~\cite{75323:2894233}\unskip~\cite{75323:2894261} emphasizes the importance of justice perceptions in understanding satisfaction and commitment among employees. Sequential linkages and the significance of evidence from procedural and interactional justice to employee turnover and commitment have been highlighted by \unskip~\cite{75323:2894234}. They reported that Person-Organization fit predicts turnover, partially mediated by job cognition and attitudes. Additionally, \unskip~\cite{75323:2894256} found that psychological well-being moderates the association between satisfaction and turnover, with a negative relationship observed for employees with lower levels of personal well-being, while this relationship was less notable for those with higher levels of well-being. Contrary to expectations, both leader influence between socialization and perceived supervisor support and social citizenship behaviors were not positively related to turnover. Another research study suggested that employees with lower levels of supervisor-rated behaviors, despite displaying organizational citizenship, have a higher likelihood of quitting their jobs\unskip~\cite{75323:2894255}.

\subsection{\textbf{Organizational Level Consequences}}There are several organizational-level consequences associated with employee turnover. Shaw and colleagues have expanded on the previous perspective of social capital, suggesting that the impact on results resulting from employee turnover is more complex than simply the loss of human capital \unskip~\cite{75323:2894271}. They have showed that the loss of organizational capital due to turnover is negatively correlated with efficiency, and the level of organizational turnover moderates this relationship. Shaw, Gupta, et al. found a curve linear relationship of workforce results and turnover in a two-unit turnover study. The negative impact was stronger when turnover was low but weakened when turnover was high.

\unskip~\cite{75323:2894236} introduced the concept of job embeddedness, which focuses on the various reasons that influence individuals to stay in their jobs. The main aspects of job embeddedness include the connections employees have with others within the organization and the community, how well they fit within the organization, and what they would have to sacrifice if they were to leave. These three dimensions of job embeddedness are referred to as links, fit, and sacrifice, and they are relevant in both the organizational and community contexts. Mitchell et al. reported the following results after conducting experiments on samples of retail and hospital employees.

Job embeddedness is a complex construct comprising six sub-dimensions, which collectively form the suggested 3 * 2 matrices. Initially, job embeddedness was calculated as an accumulated score over items for each dimension. Subsequently, accumulated job embeddedness was found to be negatively correlated with intentions to leave and predicted subsequent voluntary turnover. Moreover, job embeddedness predicted voluntary turnover even after controlling for other turnover factors such as job satisfaction, organizational commitment, and perceived alternatives.

Lee, Mitchel, Sablynski, Burton, and Holtom further expanded on research regarding job embeddedness. Their useful contribution involved disaggregating job embeddedness into two notable sub-dimensions. Utilizing a large sample of bank employees, their results showed that off-the-job embeddedness could significantly predict voluntary turnover. Additionally, the research explored potential differences between Caucasian and Hispanic groups concerning turnover and job embeddedness\unskip~\cite{75323:2894237}. Their findings indicated that job embeddedness may vary across demographic groups but does not clearly predict employee retention.

The research conducted by \unskip~\cite{75323:2894273} has revealed that organizational social tactics such as collective, fixed, and investiture methods facilitate the active integration of new employees into the organization. Their findings indicate that job embeddedness acts as a mediator in the relationship between socialization tactics and employee turnover.

Furthermore, \unskip~\cite{75323:2894270} discovered that network centrality and interpersonal citizenship behavior were negatively correlated with turnover, suggesting that employees with stronger networks and better interpersonal relationships are less likely to leave their jobs. Additionally, \unskip~\cite{75323:2894263} published findings suggesting that off-the-job embeddedness reduces turnover among women, highlighting the importance of factors outside of the workplace in influencing turnover decisions.

Moreover,\unskip~\cite{75323:2894272} undertook a reconceptualization of job embeddedness, transitioning it from a formative to a reflective construct. They investigated how general and composite forms of job embeddedness converge into Mobley Type turnover parameters. Their results showed that general job embeddedness was clearly associated with intentions to quit, searching for new job opportunities, and actual turnover. Conversely, composite job embeddedness was linked to the intention to search for and quit a job but not to actual turnover.

\unskip~\cite{75323:2894269} has endeavored to unify the parameters of job embeddedness theory and develop a comprehensive model. Their experimental findings from a study conducted at the national level, focusing on employees staying and leaving, indicated that job stayers exhibit the highest level of job embeddedness. Following them, shock-induced leavers showed the next highest level of job embeddedness, while non-shock-induced leavers showed the lowest level. Additionally, \unskip~\cite{75323:2894262} found that supportive human resource practices in organizations, such as fairness of rewards and participation in decision-making, increase perceived organizational support (POS) among employees.

Furthermore, \unskip~\cite{75323:2894204} has advocated for further research in the area of dynamic modeling of the turnover process, emphasizing how turnover unfolds over time. Many researchers have delved into dynamically modeling the turnover process, incorporating behavioral and attitudinal changes over time. For instance, researchers like Sturman and Trevor discovered that quitters' results remains relatively stable over time, whereas stayers' results shows positive trends.

\unskip~\cite{75323:2894278} has introduced a relatively unexplored job searching model, delineating three distinct job phases{\textemdash}passive planning, focused searching, and communicating with potential employers{\textemdash}and two gateways of job searching: instant job offers and financial benefits. They underscore the importance of knowledge acquisition during the job search phase, asserting that dynamic search processes enhance individuals' understanding of the employment market, especially for those contemplating quitting their current job. Moreover, they highlight that the relationship between perceived job alternatives and actual outcomes differs between individuals who intend to stay at their current organization and those who are less advanced in their job search process.

Moreover, \unskip~\cite{75323:2894264} showed that changes in effective and normative assurance are associated with variations in turnover intention over a six-month period. Changes in turnover intention over three months are related to subsequent changes over the following three months.

Lastly, researchers like \$ have conducted insightful studies comparing static turnover models, based on estimates from a single time period, with dynamic turnover models incorporating data from multiple time periods. Their findings indicate that dynamic models better fit the data. Specifically, they observed that leavers become less satisfied and committed over time, with increased withdrawal from their current job and heightened search for other job opportunities.

The above research studies have shown that we can enhance our comprehension to turnover procedure by investigating behavioral, attitudinal, contextual antecedents over time. Also, there is a need to study these factors in context with particular industry.

\subsection{\textbf{Employee Turnover In Context With IT Industry}}Some researchers have focused on studying employee turnover specifically within the IT industry, while others have applied generic theories to this context.

For example, \unskip~\cite{75323:2894240,75323:2894201,75323:2931936} conducted research on the impact of social networks and value connectivity on turnover intentions among employees in both public and private sectors. They found that clear intra-organizational social networks, characterized by good relationships and a sense of responsibility towards colleagues, reduce turnover intentions. Conversely, active external social networks increase external opportunities, making it more likely for employees to consider leaving the organization. They also proposed that person-organization fit plays a role in shaping turnover intentions, suggesting that employees with a clear fit in terms of value congruence are more likely to exhibit long-term commitment.

Similarly, \unskip~\cite{75323:2894240} identified various factors that disrupt the relationship between job satisfaction and the intention to seek employment elsewhere. Through qualitative analysis of 10 cases, they identified new disruptions, contributing to a deeper understanding of employee turnover patterns.

According to\unskip~\cite{75323:2894241}, individuals working in the IT industry operate in a dynamic environment that necessitates regular skill renewal. They argue that an individual's growth and development needs are relevant factors influencing turnover intentions. Their model suggests that the strength of development needs interacts with job satisfaction to influence turnover intentions. Additionally, factors such as the motivational aspects of the job, role conflict, and role ambiguity affect turnover intentions through their impact on job satisfaction.

Other researchers have explored the transformative effects of leadership on creativity within organizations. For instance, research by \unskip~\cite{75323:2894244} has showed that transformational leaders enhance employees' results expectations and creativity levels. They distinguish between individual, organizational, and group creativity levels and highlight the notable role of organizational climate in fostering creativity. They also find that the link between transformational leadership and creativity is moderated by organizational climate. There statistics suggest that a large number of IT professionals are expected to retire in the next decade, while fewer new workers are entering the IT field. This is expected to increase the demand for skilled personnel in the industry, making it more challenging for organizations to retain their employees. Considering this, \unskip~\cite{75323:2894245} have identified reasons that lead IT workers to change fields, proposing contextual frameworks involving variables such as work pressure, job insecurity, and burnout. They also suggest that age moderates each of these paths.

Personal desires and goals play a notable role in the IT industry. When employees' goals and desires are not met, they become dissatisfied, affecting their motivation and other factors contributing to employee turnover. Career anchors represent the personal aspirations of employees, relevant for organizations to attract and retain top talent. Career stages illustrate employees' advancement paths within an organization, reflecting their professional growth and value to the organization.

Furthermore, research by Christina Ling-Hsing Chang, Victor Chen, Gary Klein, and James J. Jiang has examined the career transitions of 10 IS employees, noting changes in career anchors throughout their careers. They found that technical proficiency and security anchors remain notable at all career stages, while factors such as geographic security, managerial competence, and self-determination become more notable in later stages.

Moreover, \unskip~\cite{75323:2894243} have highlighted that software development in organizations is often constrained by various risks at both micro and macro levels. They identify mitigation strategies proposed by numerous workers and emphasize how the organizational environment influences individuals' abilities to identify and address software-related risks. Their research identifies dimensions of organizational environment and risk within the software industry in India through factor analysis.
    
\section{Theoretical \textbf{Model And Hypothesis}}
We've constructed a theoretical framework that reflects the findings of the reviewed literature for this research endeavor.The proposed conceptual framework is in line with the objectives of our study, visually depicted in Figure 1.This research framework, illustrated graphically, outlines the independent variables: project-specific risk factors, internal and external networking, and external social network.Job satisfaction serves as a mediating variable, while perceived job alternatives function as the moderating variable for the dependent variable, turnover intentions.Throughout our investigation, we have explored several hypotheses.

\bgroup
\fixFloatSize{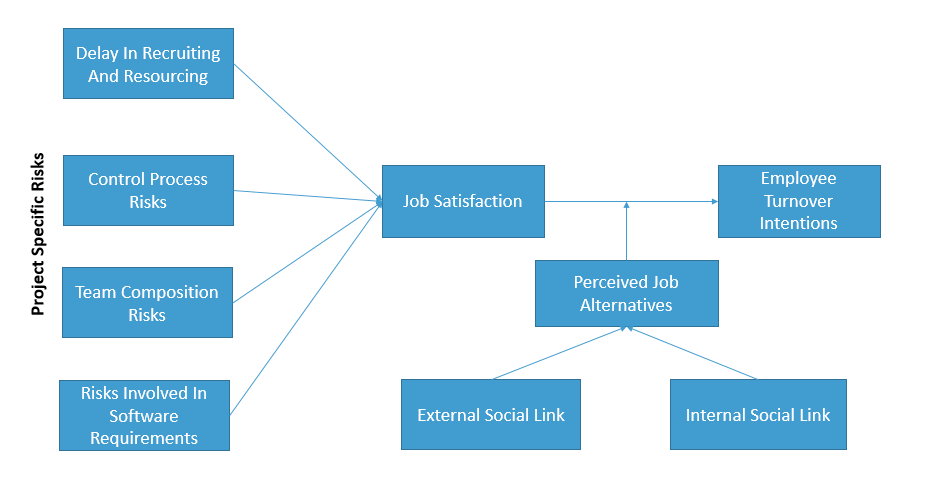}
\begin{figure*}[!htbp]
\centering \makeatletter\IfFileExists{images/11c43742-3b13-430c-aa18-384603fcb360-upaperprojectmangment.png}{\includegraphics{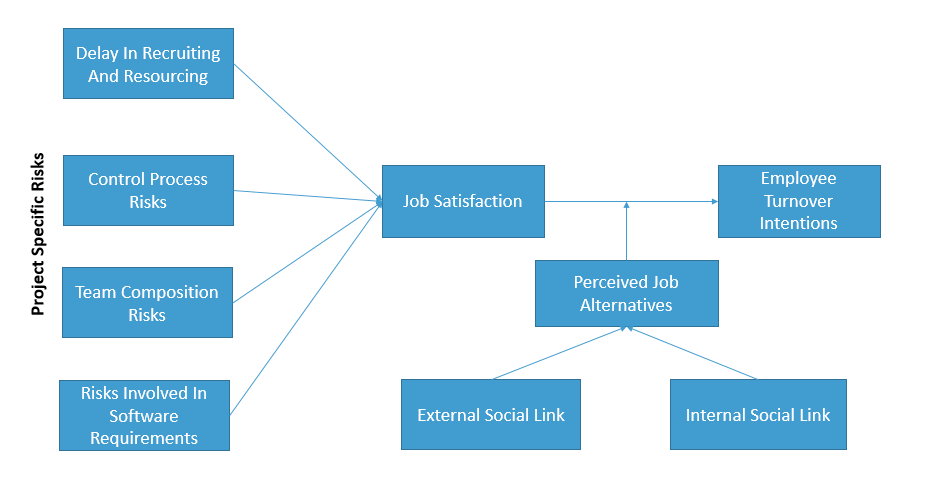}}{\includegraphics{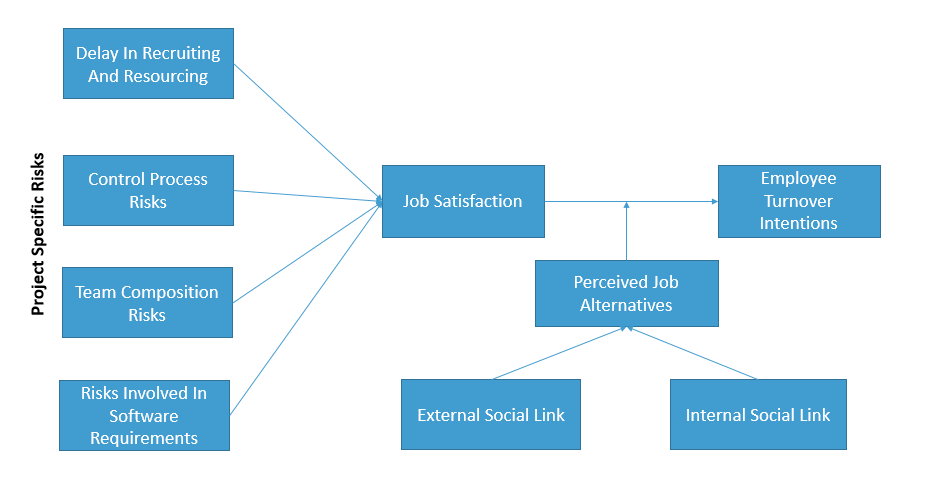}}
\makeatother 
\caption{{\textbf{ Diagram of Conceptual Model}}}
\label{figure-02194aad7a5334262283765e1d2fc7bf}
\end{figure*}
\egroup
Hypothesis 1(a): Delays in recruiting and resourcing have an impact on job satisfaction.

Hypothesis 1(b): Risks associated with control processes impact job satisfaction.

Hypothesis 1(c): Risks related to team composition affect job satisfaction.

Hypothesis 1(d): Risks associated with estimating software requirements impact job satisfaction.

Hypothesis 2: Job satisfaction is negatively associated with employee turnover intention.

Hypothesis 3: Perceived job alternatives moderate the relationship between job satisfaction and employee turnover intention.

Hypothesis 4: Both external and internal social networks positively influence perceived job alternatives.

Hypothesis 5(a): The relationship between delays in recruiting and resourcing and turnover intentions is mediated by job satisfaction.

Hypothesis 5(b): Job satisfaction mediates the relationship between risks in control processes and turnover intentions.

Hypothesis 5(c): Job satisfaction mediates the relationship between risks in team composition and turnover intention.

Hypothesis 5(d): Job satisfaction mediates the relationship between risks in estimating software requirements and turnover intention.
    
\section{Methodology}
In this section, we have presented in detail the population, sample of data, instruments of research, and methods of data collection. In this study, we have adopted a deductive research method. We have explained this in detail in the below section.

\subsection{\textbf{Data Collection Method}}

\subsection{\textbf{Sampling Definition}}A sample is a subset that captures the characteristics of a larger population \unskip~\cite{75323:2931894}. When dealing with large populations, samples used for statistical sampling must be unbiased and representative of the larger population. Defining criteria for sample size is relevant in ensuring the validity of research findings. In this study, we have explored the following criteria for determining the sample size:

\begin{itemize}
  \item \relax Employees with at least three years of experience in the software industry, either in IT-related roles or as entrepreneurs running their startups. 
  \item \relax Inclusion of project managers, team leads, and professionals actively engaged in the IT industry.
\end{itemize}

\subsection{Design Of Questionnaire}We have proposed a questionnaire based on a comprehensive literature review, selecting recent scales to aim to relevance. The questions are designed to be clear and direct, facilitating easy understanding for respondents. The questionnaire comprises the following sections:

\begin{itemize}
  \item \relax Demographic Information: This section includes general questions about respondents' demographic details. 
  \item \relax Project-Specific Risks: These questions aim to assess the impact of specific risk factors on job satisfaction and employee turnover. 
  \item \relax External and Internal Network: This section gathers information on the influence of both external and internal networks on perceived job alternatives. 
  \item \relax Perceived Job Alternatives: These questions explore how perceived job alternatives moderate the relationship between job satisfaction and turnover intention.
\end{itemize}

\subsection{Survey Procedure }Various data collection techniques exist, such as face-to-face, telephone interviews, and questionnaires. In our research, we opted to distribute a questionnaire via the web. While this method may exclude individuals without computer access, we focused on IT professionals who are likely to have computer access. Therefore, we distributed our questionnaire online throughout the IT industry in notable cities across Pakistan.

\subsection{\textbf{Data Collected}}We have collected 150 responses from various cities of Pakistan. Table 1, Table 2, Table 3, Table 4 represent our respondents' dispersion based upon Gender, age, job experience, salary, and designation within the company.
\begin{table*}[!htbp]
\caption{{Gender Distribution of Data} }
\label{table-wrap-a56aeace664f5ec4abe05d05c0d8a863}
\def\arraystretch{1}
\ignorespaces 
\centering 
\begin{tabulary}{\linewidth}{LL}
\tbltoprule 
 Gender  &
   Total Employees \\
 Male  &
   101 \\
 Female  &
   49 \\
\tblbottomrule 
\end{tabulary}\par 
\end{table*}

\begin{table*}[!htbp]
\caption{{Distribution of Employees According to Age } }
\label{table-wrap-8b2906b4fc09eb245d827b539ede79dc}
\def\arraystretch{1}
\ignorespaces 
\centering 
\begin{tabulary}{\linewidth}{LL}
\tbltoprule 
 Age  &
   Total Employees \\
 Less than 25  &
   62 \\
 25-29  &
   61 \\
 30-39  &
   21 \\
 {\textgreater} 39  &
   6 \\
\tblbottomrule 
\end{tabulary}\par 
\end{table*}

\begin{table*}[!htbp]
\caption{{\textbf{Distribution of employees according to experience}} }
\label{table-wrap-13231d8fff2ac5488f24b604a762ca8f}
\def\arraystretch{1}
\ignorespaces 
\centering 
\begin{tabulary}{\linewidth}{LL}
\tbltoprule 
 Work Experience  &
   Total Employees \\
 \textless\ 5 years  &
   73 \\
 5-10 years  &
   44 \\
 10-15 years  &
   22 \\
 16-20 years  &
   10 \\
\tblbottomrule 
\end{tabulary}\par 
\end{table*}

\begin{table*}[!htbp]
\caption{{\textbf{Distribution of employees according to salary}} }
\label{table-wrap-c420e1b7ae2049f0843897e8083c9557}
\def\arraystretch{1}
\ignorespaces 
\centering 
\begin{tabulary}{\linewidth}{LL}
\tbltoprule 
 Salary  &
   Total Employees \\
 \textless\ 25000  &
   34 \\
 25000-50000  &
   61 \\
 51000-75000  &
   26 \\
 76000-100000  &
   17 \\
 {\textgreater} 100000  &
   12 \\
\tblbottomrule 
\end{tabulary}\par 
\end{table*}

\section{ Data Analysis}
We conducted various analyses to assess our hypotheses. At first, we examined Pearson's correlation to assess the relationships between:

a) Project-specific risk factors and job satisfaction 

b) External and internal network and perceived job alternatives

c) Job satisfaction, perceived job alternatives, and employee turnover.

In the second section, we performed regression analyses to examine the impact of: 

a) Software requirement specification variability risk on job satisfaction 

b) Team composition risk on job satisfaction 

c) Control process risk on job satisfaction 

d) Dependability risk on job satisfaction 

e) Job satisfaction on employee turnover 

f) Internal network on perceived job alternatives 

g) External network on perceived job alternatives

Next, we conducted hierarchical regression to ascertain whether project-specific risk factors are fully mediated through perceived job alternatives, or if they also have a direct effect on employee turnover. Additionally, we have explored the moderating impact of perceived job alternatives between job satisfaction and employee turnover.
    
\section{\textbf{Measurement Of Variables}}
We measured all the items using a 5-point Likert scale\unskip~\cite{75323:2931936}.  The scale proposed by Arpita Sharma, and Aauushi in their research paper measures Project-specific risk factors. Similarly, the three-question scale is defined to measure Job satisfaction \unskip~\cite{75323:2894280} Price and Muller scale analyze Perceived job alternatives \unskip~\cite{75323:2894203}. External networking and social networking impact on perceived job alternatives is computed using the scale proposed by \unskip~\cite{75323:2894281} and finally employee turnover intention is measured by a scale defined by\unskip~\cite{75323:2894282}.
    
\section{\textbf{Data reliability}}
The data collected from IT professionals reflects their individual experiences in the industry, likely resulting in varied responses. To aim to the reliability of our scales, we conducted a Cronbach alpha score test\unskip~\cite{75323:2931937}. The Cronbach alpha score ranges from 0 to 1, with values exceeding 0.7 considered good. Table 5 presents the Cronbach alpha scores of our scales

\begin{table*}[!htbp]
\caption{{\textbf{Cronbach Alpha Scores of Scales}} }
\label{table-wrap-e6b5b6aa69137a4926e3206bf91b724b}
\def\arraystretch{1}
\ignorespaces 
\centering 
\begin{tabulary}{\linewidth}{LLLLLL}
\tbltoprule 
 Factor  &
   Cronbach alpha score  &
   Mean  &
   Variance  &
   Standard deviations  &
   Questions \\
 Delay in recruiting and resourcing  &
   0.79  &
   31.92  &
   30.87  &
   5.55  &
   9 \\
 Dependability risk  &
   0.753  &
   10.63  &
   5.012  &
   2.23  &
   3 \\
 Team composition risk  &
   0.756  &
   22.41  &
   16.63  &
   4.07  &
   6 \\
 Control Process Risk   &
   0.76  &
   14.46  &
   9.62  &
   3.10  &
   4 \\
 Job satisfaction   &
   0.84  &
   11.73  &
   6.82 &
   2.61  &
   3 \\
 Perceived Job Alternatives  &
   0.70  &
   6.99  &
   3.27  &
   1.80  &
   2 \\
 External Network   &
   0.90  &
   18.17  &
   25.73  &
   5.072  &
   6 \\
 Internal Network  &
   0.89  &
   21.20  &
   22.25  &
   4.717  &
   6 \\
 Turnover intention  &
   0.80  &
   7.48  &
   6.88  &
   2.62  &
   3 \\
\tblbottomrule 
\end{tabulary}\par 
\end{table*}
All of our measuring scale has Cronbach alpha score greater than 0.7 which means that they are reliable.
    
\section{\textbf{Results}}
In this section, we have conducted correlation analysis for examining the relationship between job satisfaction and employee turnover, as well as to assess the association between project-specific risks and job satisfaction.Additionally, we investigated whether perceived job alternatives moderate the relationship between job satisfaction and employee turnover.

\subsection{Correlation Analysis}Pearson correlation measures the linear relationship between two variables x and y. The value of that correlation varies between -1 and 1. The higher positive values show the positive correlation between two variables, while values closer to -1 indicate a stronger negative correlation. In our experiments, we computed the correlation between project-specific risk factors and job satisfaction. Table 6 displays the results of the correlation analysis involving project-specific risk factors, perceived job alternatives, and job satisfaction.

\begin{table*}[!htbp]
\caption{{Correlation of project specific risk factors with job satisfaction} }
\label{table-wrap-30befeaeb0af2e4869a4337f6d318aa7}
\def\arraystretch{1}
\ignorespaces 
\centering 
\begin{tabulary}{\linewidth}{LL}
\tbltoprule 
 Factors  &
   Job Satisfaction  \\
 Team composition Risk   &
   0.622* \\
 Risk Evolved in estimating software requirements  &
   0.578* \\
 Control process risk   &
   0.38 \\
 Dependability risk   &
   0.501* \\
 Perceived Job Alternatives   &
   0.488* \\
 Internal Network  &
   0.548* \\
\tblbottomrule 
\end{tabulary}\par 
\end{table*}
In our analysis, we observed that team composition risk was positively correlated/associated with job satisfaction, indicating that addressing issues related to team composition was positively correlated with job satisfaction among employees. Similarly, addressing risks associated with software requirements estimation was positively correlated with job satisfaction, contributing to higher levels of satisfaction among employees. However, the association between control process risk and job satisfaction was not statistically significant, suggesting that it did not affect job satisfaction significantly. On the other hand, the effect of dependability risk on job satisfaction was notable and positive.

Furthermore, we examined the relationship of perceived job alternatives on job satisfaction and found a positive relationship between the two variables. Table 7 illustrates the relationship between job satisfaction, perceived job alternatives, and turnover intentions.

\begin{table*}[!htbp]
\caption{{\textbf{Correlation of JS and PJA with TI}} }
\label{table-wrap-9744904eb002b284f85596adf6b72754}
\def\arraystretch{1}
\ignorespaces 
\centering 
\begin{tabulary}{\linewidth}{LL}
\tbltoprule 
 Factors   &
   Turnover Intentions \\
 Job satisfaction  &
   -0.560* \\
 Perceived Job Alternatives  &
   0.16 \\
\tblbottomrule 
\end{tabulary}\par 
\end{table*}
Job satisfaction was associated with lower turnover intentions leading to fewer turnover intentions when people are happier with their work. Similarly,  perceived job opportunities have no statistically significant relationship with turnover intentions. In the final set of experiments of Pearson's correlation, we computed the relationship between internal and external networking and perceived job alternatives. Table 8 shows our results.
\begin{table*}[!htbp]
\caption{{Correlation of IN and EN with PJA} }
\label{table-wrap-5cd4496dc0a16e70dcb5e8b7251d932f}
\def\arraystretch{1}
\ignorespaces 
\centering 
\begin{tabulary}{\linewidth}{LL}
\tbltoprule 
 Factors   &
   Perceived Job Alternatives \\
 Internal networking  &
   0.200 \\
 External networking  &
   0.24 \\
\tblbottomrule 
\end{tabulary}\par 
\end{table*}

\subsection{\textbf{Direct Regression Analysis}}In our regression analysis, we investigated the relationships among project-specific risk factors, job satisfaction, turnover intentions, perceived job alternatives, and external and internal networking. Turnover intention was considered the dependent variable, while other parameters served as independent variables. Control variables such as Gender, salary, age, and job experience were included, and the analysis proceeded in two steps.

Initially, control values were entered into the model, revealing that they did not significantly affect job satisfaction, with a p-value higher than 0.05. The \(R^2\) value of control variables indicated they accounted for a 4.1\% variance in job satisfaction. Next, we added the team composition risk factor to the model. In the presence of team composition risk, control variables became significantly related to job satisfaction, explaining an additional 47.2\% variance in job satisfaction. Reductions in team composition risks was significantly positively associated with job satisfaction, with a coefficient (B) of 0.622 and p-value \textless\ 0.05. Overall, the analysis was consistent with the hypotheses 1(a), 1(c), and 1(d), indicating that team composition risks, dependability risks, and risks involved in estimating software requirements had notable effects on job satisfaction. However, control process risks did not significantly affect job satisfaction. \mbox{}\protect\newline Control process risk did not show notable correlation with job satisfaction (Beta = 0.46, p {\textgreater} 0.05), suggesting that control process risks are not significantly related to job satisfaction. In the case of dependability risk (fourth column of the table), none of the control variables showed a notable association with job satisfaction in the presence of dependability risk. However, reducing dependability risk was positively associated with job satisfaction (Beta = 0.499, p \textless\ 0.05). Dependability risks explained an additional 35.9\% variance in job satisfaction. \mbox{}\protect\newline When considering risks involved in estimating software requirements (fifth column of the table), control variables did not show notable correlation with job satisfaction. However, risks involved in estimating software requirements had a positive influence on job satisfaction (Beta = 0.578, p \textless\ 0.05). These risks explained 42.8\% of the variance in job satisfaction. \mbox{}\protect\newline Thus, hypotheses 1(a), 1(c), and 1(d) are supported, indicating that team composition risks, dependability risks, and risks involved in estimating software requirements significantly impact job satisfaction. However, hypothesis 1(b) is not supported, suggesting that control process risks do not have a notable effect on job satisfaction. \mbox{}\protect\newline Furthermore, job satisfaction was found to have a notable negative relationship with turnover intention (Beta = -0.560, p \textless\ 0.01), explaining 42\% of the variance in employee turnover intention. Moreover, perceived job alternatives were positively associated with turnover intentions, accounting for an additional 51\% of the variance in turnover intentions. The beta coefficient value of perceived job alternatives was 0.69 (p < 0.01) in the regression model (Table 10); however, Table 7 shows no significant bivariate correlation between perceived job alternatives and turnover intentions (r = 0.16), so this result should be interpreted as conditional on the regression model.

To evaluate hypotheses 4(a) and 4(b), we regressed external networking and internal networking against perceived job alternatives. Internal networking did not influence perceived job alternatives significantly (p {\textgreater} 0.05), with an \(R^2\) value of only 1.4\%, indicating that internal networking explains very little variance in perceived job alternatives. Similarly, external social networking did not have a notable influence on perceived job alternatives, with a low \(R^2\) value suggesting minimal impact on employee turnover intentions.

Table 9, Table 10, Table 11 represents the details of our results.
\begin{table*}[!htbp]
\caption{{\textbf{Direct Regression Analysis of Project Specific Risk factors}} }
\label{table-wrap-4c7798b7dbda1b73ce39d4816536c193}
\def\arraystretch{1}
\ignorespaces 
\centering 
\begin{tabulary}{\linewidth}{LLLLLL}
\tbltoprule 
 Variables   &
  \multicolumn{5}{p{\dimexpr(\mcWidth{2}+\mcWidth{3}+\mcWidth{4}+\mcWidth{5}+\mcWidth{6})}}{ Job satisfaction }\\
 Column No  &
   1 Beta  &
   2 Beta  &
   3  Beta  &
   4 Beta  &
   5 Beta \\
 Age  &
   0.14  &
   0.09  &
   0.07  &
   0.15  &
   0.06 \\
 Gender  &
   0.07  &
   0.1  &
   0.17  &
   0.09  &
   0.10 \\
 Salary Range: {\textless}25000  &
   0.18  &
   0.12  &
   0.11  &
   0.19  &
   0.16 \\
 25000 {\textendash} 50000  &
   0.21  &
   0.18  &
   0.13  &
   0.15  &
   0.19 \\
 51000 - 75000  &
   0.23  &
   0.11  &
   0.16  &
   0.21  &
   0.20 \\
 76000 - 100000  &
   0.27  &
   0.28  &
   0.222  &
   0.23  &
   0.18 \\
 {\textgreater}100000  &
   0.34  &
   0.30  &
   0.33  &
   0.27  &
   0.26 \\
 Work Experience {\textless}5 years   &
   0.16  &
   0.13  &
   0.21  &
   0.18  &
   0.17 \\
 5-10 years  &
   0.22  &
   0.21  &
   0.14  &
   0.21  &
   0.18 \\
 10-15 years  &
   0.29  &
   0.27  &
   0.11  &
   0.17  &
   0.23 \\
 16-20 years  &
   0.31  &
   0.30  &
   0.16  &
   0.19  &
   0.21 \\
 TCR  &
   -  &
   0.622*  &
   -  &
   -  &
   - \\
 CPR  &
   -  &
   -  &
   0.46  &
   -  &
   - \\
 DR  &
   -  &
   -  &
   -  &
   0.499*  &
   - \\
 RIESR  &
   -  &
   -  &
   -  &
   -  &
   0.578* \\
 \(R^2\)  &
   -  &
   0.481  &
   0.073  &
   0.362  &
   0.584 \\
 $(\Delta R^2)$  &
   -  &
   0.472  &
   0.087  &
   0.359  &
   0.578 \\
\tblbottomrule 
\end{tabulary}\par 
\end{table*}

\begin{table*}[!htbp]
\caption{{Direct Regression Analysis of PJA and JS} }
\label{table-wrap-e45a22323bc2b4408f38c65d8e7be0c0}
\def\arraystretch{1}
\ignorespaces 
\centering 
\begin{tabulary}{\linewidth}{LLL}
\tbltoprule 
 Variables  &
  \multicolumn{2}{p{\dimexpr(\mcWidth{2}+\mcWidth{3})}}{ Turnover Intentions }\\
 Column No  &
   1 Beta  &
   2 Beta \\
 Age  &
   0.10  &
   0.13 \\
 Gender  &
   0.07  &
   0.09 \\
 Salary Range: {\textless}25000  &
   0.18  &
   0.19 \\
 25000 {\textendash} 50000  &
   0.20  &
   0.21 \\
 51000 - 75000  &
   0.22  &
   0.23 \\
 76000 - 100000  &
   0.27  &
   0.25 \\
 {\textgreater}100000  &
   0.29  &
   0.26 \\
 Work Experience {\textless}5 years   &
   0.11  &
   0.17 \\
 5-10 years  &
   0.12  &
   0.21 \\
 10-15 years  &
   0.14  &
   0.22 \\
 16-20 years  &
   0.13  &
   0.27 \\
 JS  &
   -0.560*  &
   - \\
 PJA  &
   -  &
   0.69* \\
 \(R^2\)  &
   0.41  &
   0.55 \\
 $(\Delta R^2)$  &
   0.42  &
   0.51 \\
\tblbottomrule 
\end{tabulary}\par 
\end{table*}

\begin{table*}[!htbp]
\caption{{Direct Regression Analysis of EN and IN} }
\label{table-wrap-57cc74f1f81cbc6ccdbd351845f2f21f}
\def\arraystretch{1}
\ignorespaces 
\centering 
\begin{tabulary}{\linewidth}{LLL}
\tbltoprule 
 Variables  &
  \multicolumn{2}{p{\dimexpr(\mcWidth{2}+\mcWidth{3})}}{ PJA }\\
 Column No  &
   1 Beta  &
   2 Beta \\
 External Networking   &
   0.16  &
   - \\
 Internal Networking  &
   -  &
   0.21 \\
 \(R^2\)  &
   0.018  &
   0.026 \\
 $(\Delta R^2)$  &
   0.014  &
   0.027 \\
\tblbottomrule 
\end{tabulary}\par 
\end{table*}

\subsection{\textbf{Moderating Effect of perceived Job Alternatives:}}To evaluate our hypothesis (2), we investigated whether other job opportunities moderate the association between job satisfaction and employee turnover intentions. We conducted a hierarchical moderated regression for this purpose.

In step 1 of our experiment, we introduced job satisfaction as the independent variable. Job satisfaction exhibited a notable negative effect on turnover intention (B = -0.565, p \textless\ 0.01), explaining 43\% of the variance in employee turnover intention.

In step 2, we added perceived job alternatives to the model. The regression coefficient for perceived job alternatives was found to be B = 0.714, indicating a notable positive impact on turnover intention. Perceived job alternatives explained an additional 52\% of the variance in turnover intention. Consequently, the predictive capacity of our model improved with the inclusion of perceived job alternatives. Together, job satisfaction and perceived job alternatives accounted for 58\% of the variance in turnover intentions. In step 3, we introduced an interaction term into the model by multiplying job satisfaction and perceived job alternatives. The purpose was to assess whether perceived job alternatives moderate the relationship between job satisfaction and turnover intention. However, the results showed that the interaction term had a not statistically significant impact (B = 0.095, p {\textgreater} 0.05), indicating that perceived job alternatives do not moderate the relationship between job satisfaction and employee turnover. Therefore, hypothesis (3) of our study is not supported.

Table 12 presents the results of the hierarchical moderated regression analysis.

\begin{table*}[!htbp]
\caption{{Hierarchical Moderated Regression Results} }
\label{table-wrap-1e90685d02c403818191aba3c94b9be3}
\def\arraystretch{1}
\ignorespaces 
\centering 
\begin{tabulary}{\linewidth}{LLLL}
\tbltoprule 
 Variables  &
   STEP1  &
   STEP2  &
   STEP3 \\
 JS  &
   -.565**  &
   -0.578**  &
   -0.618* \\
 PJA  &
   -  &
   0.714**  &
   0.456* \\
 Interaction Term (JS * PJA)  &
   &
   &
   0.095 \\
 \(R^2\)  &
   -.565  &
   0.65  &
   0.387 \\
\tblbottomrule 
\end{tabulary}\par 
\end{table*}
** denotes notable two tail test where p\textless\ 0.01 * denotes the 1 tail test where p \textless\ 0.05

\subsection{\textbf{Mediation Test }}To evaluate hypothesis (3), we analyzed whether job satisfaction fully or partially mediates the association between project-specific risk factors and turnover intention. In step 1 of our analysis, control process risk was found to have no notable direct association with turnover intention (beta = -0.21, p {\textgreater} 0.05). In step 2, control process risk was regressed on job satisfaction, revealing a not statistically significant relationship (beta = 0.19, p {\textgreater} 0.01). However, in step 3, job satisfaction was negatively associated with turnover intentions (beta = -0.561, p \textless\ 0.01).

In the final step, where job satisfaction was included as a control variable, the relationship between control process risk and turnover intention remained insignificant. These findings suggest that job satisfaction does not fully mediate the association between control process risk and turnover intention.

Table 13 presents the detailed results of our analysis

\begin{table*}[!htbp]
\caption{{\textbf{Results of mediated Regression for CPR}} }
\label{table-wrap-0295c650f332f823a3be74a2d5d6641d}
\def\arraystretch{1}
\ignorespaces 
\centering 
\begin{tabulary}{\linewidth}{LLLLL}
\tbltoprule 
 Variables  &
   Step1  &
   Step2  &
   Step3  &
   Step4 \\
 CPR  &
   -0.21  &
   0.19  &
   -  &
   0.18 \\
 JS  &
   -  &
   -  &
   -0.561**  &
   -0.475** \\
 \(R^2\)  &
   0.017  &
   0.022  &
   0.48  &
   0.114 \\
\tblbottomrule 
\end{tabulary}\par 
\end{table*}
Dependability risk has a notable impact on turnover intentions (Beta = -0.501, p \textless\ 0.05). In step 2, we examined the relationship between Dependability risk and job satisfaction, finding that Dependability risks was positively correlated with job satisfaction (Beta = 0.498, p \textless\ 0.01). Step 3 revealed that job satisfaction significantly affects turnover intentions (beta = -0.582, p \textless\ 0.01). In step four of the experiment, job satisfaction was used as a control variable. Multiple regression analysis showed that both Dependability risks (beta = -0.452, p \textless\ 0.01) and job satisfaction (beta = -0.356, p \textless\ 0.01) impact turnover intention. This suggests that job satisfaction partially mediates the relationship between Dependability risk and turnover intentions.

Table 14 provides detailed results of our experiment.

\begin{table*}[!htbp]
\caption{{\textbf{Results of mediated Regression for DR}} }
\label{table-wrap-d96dd90bca6f365093cc61786f373948}
\def\arraystretch{1}
\ignorespaces 
\centering 
\begin{tabulary}{\linewidth}{LLLLL}
\tbltoprule 
 Variables  &
   Step1  &
   Step2  &
   Step3  &
   Step4 \\
 DR  &
   -0.501*  &
   0.498**  &
   -  &
   -0.452** \\
 JS  &
   -  &
   -  &
   -0.582**  &
   -0.356** \\
 \(R^2\)  &
   0.369  &
   0.378  &
   0.49  &
   0.46 \\
\tblbottomrule 
\end{tabulary}\par 
\end{table*}
Risks involved in estimating software requirements (RIESR) also has a notable impact on turnover intentions, with a Beta value of 0.578 and p \textless\ 0.01. In step 2 of our experiment, we included job satisfaction in the model to assess the association between RIESR and job satisfaction. Similarly, RIESR is significantly related to job satisfaction (Beta = 0.547, p \textless\ 0.01). In step 3, we examined the relationship between job satisfaction and turnover intentions. Job satisfaction was found to be negatively related to turnover intentions (Beta = -0.596, p \textless\ 0.01). In step 4, job satisfaction was used as a control variable. Multiple regression analysis revealed that both RIESR (Beta = 0.571, p \textless\ 0.01) and job satisfaction (Beta = 0.456, p \textless\ 0.01) impact turnover. This indicates that satisfaction with a job partially mediates the association between RIESR and turnover intentions.

Table 15 provides detailed results of the mediated regression for RIESR.

\begin{table*}[!htbp]
\caption{{\textbf{Results of Mediated Regression for RIESR}} }
\label{table-wrap-fbf01a04ca87bc684ca36a22170dbbe5}
\def\arraystretch{1}
\ignorespaces 
\centering 
\begin{tabulary}{\linewidth}{LLLLL}
\tbltoprule 
 Variables  &
   Step1  &
   Step2  &
   Step3  &
   Step4 \\
 RIESR  &
   -0.578*  &
   0.547**  &
   -  &
   -0.571** \\
 JS  &
   -  &
   -  &
   -0.596**  &
   -0.456** \\
 \(R^2\)  &
   0.369  &
   0.378  &
   0.49  &
   0.46 \\
\tblbottomrule 
\end{tabulary}\par 
\end{table*}
In the final experiment, we examined whether Team Risks mediate through job satisfaction and significantly influence employee turnover. In step 1, we assessed the influence of Team Risks on turnover intentions. Team risk was found to have a notable impact on turnover intentions (Beta = -0.512, p \textless\ 0.01). In step 2, we introduced job satisfaction into the model. Team risks were significantly related to job satisfaction (Beta = 0.642, p \textless\ 0.01). In step 3, we analyzed the influence of job satisfaction on turnover intentions (Beta = 0.578, p \textless\ 0.01). Finally, in step 4, job satisfaction was included as a control variable. Multiple regression analysis revealed that both team composition risk (Beta = 0.593, p \textless\ 0.01) and job satisfaction (Beta = 0.481, p \textless\ 0.01) significantly influence turnover intentions. This suggests that job satisfaction partially mediates the association between team composition risk and turnover intentions.
    
\section{\textbf{Conclusions}}
The research study delves into the impact of project-specific risks on turnover intentions, the role of external and internal networking in perceived job alternatives in Pakistan's IT industry, and the development of a statistical analysis to examine employee turnover intentions. This exploration is particularly context-specific in the context of the IT industry in Pakistan, which plays a relevant role in driving the country's economy and creating employment opportunities. Given that IT professionals often juggle multiple projects simultaneously, they encounter various project-specific risks. Understanding these risks and their effects is vital for project success and mitigating turnover intentions. This study aimed to investigate the influence of project risks on turnover intentions, mediated by job satisfaction, as well as to explore the moderating role of perceived job alternatives (PJA). Additionally, the study sought to examine the impact of external and internal networking on PJA, alongside the main objective of developing an analytical framework for employee turnover intentions.

To gather data, we distributed a questionnaire to numerous companies across Pakistan, resulting in responses from 150 participants. Our analyses encompassed various statistical techniques, including regression analysis, step-wise linear regression, and correlation analysis, to evaluate our hypotheses. Our findings revealed that minimizing project-specific risks is important for reducing turnover intentions. Specifically, all project-specific risk factors, with the exception of control process risk, were associated with turnover intention through job satisfaction and influenced employee turnover. Additionally, while perceived job alternatives did not moderate the relationship between job satisfaction and turnover intention, they independently influenced turnover intention. Similarly, our analyses indicated that external and internal networking did not directly impact turnover intention. In summary, this study sheds light on critical factors influencing turnover intentions in Pakistan's IT industry and underscores the importance of managing project-specific risks and perceived job alternatives to mitigate turnover.
    
\section{\textbf{Future Work}}
In the future, we plan to expand our research to explore project-specific risks across diverse industries, such as construction companies. By investigating the factors contributing to project-specific risks in various sectors, we aim to gain a comprehensive understanding of their influence on employee turnover intentions. Additionally, we intend to delve deeper into the determinants of employee turnover intentions, examining a wide range of influencing factors. This comprehensive approach will enable us to uncover detailed insights into the drivers of turnover intentions and their implications for organizational management. Moreover, we recognize the importance of considering cultural differences in our research. We aim to conduct cross-cultural analyses to explore how cultural norms and values shape employee turnover intentions. By collecting data from different cultural contexts, we can better understand the variations in turnover intentions across diverse populations.

Furthermore, we plan to evaluate the outcomes of organizational initiatives and activities in mitigating employee turnover intentions. By assessing the impact of various organizational interventions, such as training programs, leadership initiatives, and employee engagement strategies, we can identify effective measures for reducing turnover and retaining skilled talent. Overall, our future research endeavors will contribute to the development of evidence-based strategies for minimizing employee turnover intentions and informing organizational outcomes, particularly in dynamic industries like IT where talent retention is relevant.

\section{Ethical considerations}
Participation was voluntary. No direct identifiers were collected. Responses were used only for research and reported in aggregate.

\bibliographystyle{apalike}
\bibliography{article}

\section{\textbf{Appendix}}

\subsection{\textbf{Questionnaire}}
\noindent\textbf{How much is your experience (years) in industry?} \textsuperscript{*}
\begin{itemize}[leftmargin=2em,itemsep=0.25em]
  \item  < 5 years
  \item  5--10 years
  \item  11--15 years
  \item  16--20 years
  \item  > 20 years
\end{itemize}

\medskip

\noindent\textbf{What is your salary range?} \textsuperscript{*}
\begin{itemize}[leftmargin=2em,itemsep=0.25em]
  \item  < 25{,}000
  \item  25{,}000--50{,}000
  \item  51{,}000--75{,}000
  \item  76{,}000--100{,}000
  \item  > 100{,}000
\end{itemize}

\medskip

\noindent\textbf{What is the company legal status?} \textsuperscript{*}
\begin{itemize}[leftmargin=2em,itemsep=0.25em]
  \item  Public firm
  \item  Private firm
\end{itemize}
\noindent\textbf{How many total employees your company has?} \textsuperscript{*}\\[0.5ex]

\medskip

\noindent\textbf{Is your company a startup venture?} \textsuperscript{*}
\begin{itemize}[leftmargin=2em,itemsep=0.25em]
  \item  Yes
  \item  No
\end{itemize}

\medskip

\noindent\textbf{Age} \textsuperscript{*}
\begin{itemize}[leftmargin=2em,itemsep=0.25em]
  \item  Less than 25
  \item  25--29
  \item  30--39
  \item  40--49
  \item  50 or above
\end{itemize}



\noindent\textbf{Please Rate the influence of project specific risks factor on your job satisfaction?} \textsuperscript{*}\\
\emph{a) Risks involved in Estimating software requirements}

\begin{center}
\begingroup
\setlength{\tabcolsep}{2pt}     
\renewcommand{\arraystretch}{1.1}
\small                           

\begin{tabular}{|p{0.35\textwidth}|*{5}{>{\centering\arraybackslash}p{0.125\textwidth}|}}
\hline
 & \textbf{Unimportant} & \textbf{Little Important} & \textbf{Moderately Important} & \textbf{Important} & \textbf{Very Important} \\
\hline
Delay in recruiting and resourcing & $\circ$ & $\circ$ & $\circ$ & $\circ$ & $\circ$ \\ \hline
Less or no experienced in similar project & $\circ$ & $\circ$ & $\circ$ & $\circ$ & $\circ$ \\ \hline
Estimation errors & $\circ$ & $\circ$ & $\circ$ & $\circ$ & $\circ$ \\ \hline
Inaccurate requirement analysis & $\circ$ & $\circ$ & $\circ$ & $\circ$ & $\circ$ \\ \hline
Miscommunication of requirements & $\circ$ & $\circ$ & $\circ$ & $\circ$ & $\circ$ \\ \hline
Conflicting and continuous requirements & $\circ$ & $\circ$ & $\circ$ & $\circ$ & $\circ$ \\ \hline
Language and regional differences with clients & $\circ$ & $\circ$ & $\circ$ & $\circ$ & $\circ$ \\ \hline
Lack of client ownership and owner stability & $\circ$ & $\circ$ & $\circ$ & $\circ$ & $\circ$ \\ \hline
Inaccurate cost measurement & $\circ$ & $\circ$ & $\circ$ & $\circ$ & $\circ$ \\ \hline
\end{tabular}
\endgroup
\end{center}

\noindent\textbf{b) Team Composition Risk} \textsuperscript{*}

\begin{center}
\renewcommand{\arraystretch}{1.2}
\setlength{\tabcolsep}{3pt} 

\begin{tabular}{|>{\raggedright\arraybackslash}p{0.35\textwidth}|*{5}{>{\centering\arraybackslash}p{0.125\textwidth}|}}
\hline
 & \textbf{Un important} & \textbf{Little Important} & \textbf{Moderately Important} & \textbf{Important} & \textbf{Very Important} \\
\hline
Working with inexperienced team        & $\circ$ & $\circ$ & $\circ$ & $\circ$ & $\circ$ \\ \hline
Team diversity                         & $\circ$ & $\circ$ & $\circ$ & $\circ$ & $\circ$ \\ \hline
Lack of availability of domain expert  & $\circ$ & $\circ$ & $\circ$ & $\circ$ & $\circ$ \\ \hline
Lack of commitment from project team   & $\circ$ & $\circ$ & $\circ$ & $\circ$ & $\circ$ \\ \hline
Lack of top management support         & $\circ$ & $\circ$ & $\circ$ & $\circ$ & $\circ$ \\ \hline
Low morale of the team                 & $\circ$ & $\circ$ & $\circ$ & $\circ$ & $\circ$ \\ \hline
\end{tabular}
\end{center}
\vspace{1\baselineskip}
\noindent\textbf{Please select the impact of branding employment and Peer perception regarding work on job satisfaction?} \textsuperscript{*}\\
\emph{Branding Employment means how people see the worth of job oppourtunity in the organization. 
Peer Perception Regarding Work means how your friends and other peers value you based upon your work}

\begin{center}
\small
\setlength{\tabcolsep}{3pt}
\renewcommand{\arraystretch}{1.15}

\begin{tabular}{|>{\raggedright\arraybackslash}p{0.39\textwidth}|*{5}{>{\centering\arraybackslash}p{0.11\textwidth}|}}
\hline
 & \textbf{Strongly Disagree} & \textbf{Disagree} & \textbf{Neutral} & \textbf{Agree} & \textbf{Strongly Agree} \\
\hline
External friends (like university friends and other friends of your field which are not working in your organization) finds your work interesting and innovative & $\circ$ & $\circ$ & $\circ$ & $\circ$ & $\circ$ \\ \hline
Internal peers sees tasks assigned to you interesting and innovative & $\circ$ & $\circ$ & $\circ$ & $\circ$ & $\circ$ \\ \hline
You have a notable position in the community based upon the perception regarding your work & $\circ$ & $\circ$ & $\circ$ & $\circ$ & $\circ$ \\ \hline
Work experience of your organization is given more preference as compared to work experience of others & $\circ$ & $\circ$ & $\circ$ & $\circ$ & $\circ$ \\ \hline
Your company creates a sense of urgency and excitement about working at the company & $\circ$ & $\circ$ & $\circ$ & $\circ$ & $\circ$ \\ \hline
Your organization engages mind, heart and dreams & $\circ$ & $\circ$ & $\circ$ & $\circ$ & $\circ$ \\ \hline
Your organization provides a clear and compelling reason to work & $\circ$ & $\circ$ & $\circ$ & $\circ$ & $\circ$ \\ \hline
Your company is consistent with what employee believes about working at the organization. & $\circ$ & $\circ$ & $\circ$ & $\circ$ & $\circ$ \\ \hline
Your company evokes feeling of fun, prestige, challenge and rewards & $\circ$ & $\circ$ & $\circ$ & $\circ$ & $\circ$ \\ \hline
\end{tabular}
\end{center}
\vspace{1\baselineskip}
\noindent\textbf{Please choose the importance of organization adaptation to changing technology and career opportunities on the job satisfaction} \textsuperscript{*}\\
\emph{a) Career opportunities}

\begin{center}
\small
\setlength{\tabcolsep}{3pt}
\renewcommand{\arraystretch}{1.15}

\begin{tabular}{|>{\raggedright\arraybackslash}p{0.40\textwidth}|*{5}{>{\centering\arraybackslash}p{0.11\textwidth}|}}
\hline
 & \textbf{Strongly Disagree} & \textbf{Disagree} & \textbf{Neutral} & \textbf{Agree} & \textbf{Strongly Agree} \\
\hline
Adequate work fulfillment                         & $\circ$ & $\circ$ & $\circ$ & $\circ$ & $\circ$ \\ \hline
Work useful enough                                & $\circ$ & $\circ$ & $\circ$ & $\circ$ & $\circ$ \\ \hline
Good use of your skills                           & $\circ$ & $\circ$ & $\circ$ & $\circ$ & $\circ$ \\ \hline
Chances to improve skills                         & $\circ$ & $\circ$ & $\circ$ & $\circ$ & $\circ$ \\ \hline
Promotional opportunities                         & $\circ$ & $\circ$ & $\circ$ & $\circ$ & $\circ$ \\ \hline
Lateral transfer opportunities                    & $\circ$ & $\circ$ & $\circ$ & $\circ$ & $\circ$ \\ \hline
Independence on the job                           & $\circ$ & $\circ$ & $\circ$ & $\circ$ & $\circ$ \\ \hline
Providing good quality on-the-job training        & $\circ$ & $\circ$ & $\circ$ & $\circ$ & $\circ$ \\ \hline
Good chance of getting formal training            & $\circ$ & $\circ$ & $\circ$ & $\circ$ & $\circ$ \\ \hline
Good quality formal training                      & $\circ$ & $\circ$ & $\circ$ & $\circ$ & $\circ$ \\ \hline
\end{tabular}
\end{center}
\vspace{1\baselineskip}
\noindent\textbf{b) Organization adaptation to changing technology}

\begin{center}
\small
\setlength{\tabcolsep}{2pt}
\renewcommand{\arraystretch}{1.15}

\begin{tabular}{|>{\raggedright\arraybackslash}p{0.48\textwidth}|*{5}{>{\centering\arraybackslash}p{0.10\textwidth}|}}
\hline
 & \textbf{Strongly Disagree} & \textbf{Disagree} & \textbf{Neutral} & \textbf{Agree} & \textbf{Strongly Agree} \\
\hline
Adaptation of cutting edge commonly used technology for company products and services & $\circ$ & $\circ$ & $\circ$ & $\circ$ & $\circ$ \\ \hline
Providing adequate training programs for commonly used technology used in company line of business & $\circ$ & $\circ$ & $\circ$ & $\circ$ & $\circ$ \\ \hline
Adaptation of latest principles, mechanisms and techniques for improving productivity and quality & $\circ$ & $\circ$ & $\circ$ & $\circ$ & $\circ$ \\ \hline
Advance planning for adaptation of futuristic technology as compared to other organizations & $\circ$ & $\circ$ & $\circ$ & $\circ$ & $\circ$ \\ \hline
Adequate financial resources required for quick change to new technology & $\circ$ & $\circ$ & $\circ$ & $\circ$ & $\circ$ \\ \hline
\end{tabular}
\end{center}

\vspace{1\baselineskip}
\noindent\textbf{Please choose the importance and rate the influence of external social network, external and internal networking on perceived job alternatives}\\
\emph{External Social Network:}

\begin{center}
\setlength{\tabcolsep}{3pt}
\renewcommand{\arraystretch}{1.2}

\begin{tabular}{|>{\raggedright\arraybackslash}p{0.73\textwidth}|>{\centering\arraybackslash}p{0.12\textwidth}|>{\centering\arraybackslash}p{0.12\textwidth}|}
\hline
 & \textbf{Yes} & \textbf{No} \\
\hline
a) Are you a member of any professional association? \textsuperscript{*} & $\circ$ & $\circ$ \\ \hline
b) I regularly attend professional association meetings at the local level \textsuperscript{*} & $\circ$ & $\circ$ \\ \hline
c) I systematically read professional journals \textsuperscript{*} & $\circ$ & $\circ$ \\ \hline
d) Have you joined professional network site (e.g., LinkedIn)? \textsuperscript{*} & $\circ$ & $\circ$ \\ \hline
\end{tabular}
\end{center}

\begin{center}
\setlength{\tabcolsep}{3pt}
\renewcommand{\arraystretch}{1.2}

\begin{tabular}{|>{\raggedright\arraybackslash}p{0.73\textwidth}|>{\centering\arraybackslash}p{0.13\textwidth}|>{\centering\arraybackslash}p{0.13\textwidth}|}
\hline
 & \textbf{Yes} & \textbf{No} \\
\hline
e) Do you regularly maintain your profile on professional network site (e.g., LinkedIn)? \textsuperscript{*} & $\circ$ & $\circ$ \\
\hline
\end{tabular}
\end{center}
\vspace{1\baselineskip}
\noindent\textbf{External networking} \textsuperscript{*}

\begin{center}
\small
\setlength{\tabcolsep}{2.5pt}
\renewcommand{\arraystretch}{1.15}

\begin{tabular}{|>{\raggedright\arraybackslash}p{0.52\textwidth}|*{5}{>{\centering\arraybackslash}p{0.09\textwidth}|}}
\hline
 & \textbf{Strongly Disagree} & \textbf{Disagree} & \textbf{Neutral} & \textbf{Agree} & \textbf{Strongly Agree} \\
\hline
I spend a lot of time and effort at work networking outside my organization & $\circ$ & $\circ$ & $\circ$ & $\circ$ & $\circ$ \\ \hline
I am good at building relationships with influential people outside my organization & $\circ$ & $\circ$ & $\circ$ & $\circ$ & $\circ$ \\ \hline
I know a lot of important people and am well connected outside my organization & $\circ$ & $\circ$ & $\circ$ & $\circ$ & $\circ$ \\ \hline
I spend a lot of time developing connections with others outside my organization & $\circ$ & $\circ$ & $\circ$ & $\circ$ & $\circ$ \\ \hline
I am good at using my connections and network outside my organization to make things happen for my career & $\circ$ & $\circ$ & $\circ$ & $\circ$ & $\circ$ \\ \hline
I have developed a large network of colleagues and associates outside my organization that I can call on for support when I really need to get things done & $\circ$ & $\circ$ & $\circ$ & $\circ$ & $\circ$ \\ \hline
\end{tabular}
\end{center}
\vspace{1\baselineskip}
\noindent\textbf{Internal networking} \textsuperscript{*}

\begin{center}
\small
\setlength{\tabcolsep}{2.5pt}
\renewcommand{\arraystretch}{1.15}

\begin{tabular}{|>{\raggedright\arraybackslash}p{0.52\textwidth}|*{5}{>{\centering\arraybackslash}p{0.09\textwidth}|}}
\hline
 & \textbf{Strongly Disagree} & \textbf{Disagree} & \textbf{Neutral} & \textbf{Agree} & \textbf{Strongly Agree} \\
\hline
I spend a lot of time and effort networking with others within my organization & $\circ$ & $\circ$ & $\circ$ & $\circ$ & $\circ$ \\ \hline
I am good at building relationships with influential people within my organization & $\circ$ & $\circ$ & $\circ$ & $\circ$ & $\circ$ \\ \hline
I know a lot of important people and am well connected within my organization & $\circ$ & $\circ$ & $\circ$ & $\circ$ & $\circ$ \\ \hline
I spend a lot of time developing connections with others within my organization & $\circ$ & $\circ$ & $\circ$ & $\circ$ & $\circ$ \\ \hline
I am good at using my connections and network within my organization to make things happen at work & $\circ$ & $\circ$ & $\circ$ & $\circ$ & $\circ$ \\ \hline
I have developed a large network of colleagues and associates within my organization whom I can call on for support when I really need to get things done & $\circ$ & $\circ$ & $\circ$ & $\circ$ & $\circ$ \\ \hline
\end{tabular}
\end{center}

\vspace{1\baselineskip}
\noindent\textbf{Please choose how perceived job alternatives mediates or moderates the relationship between job satisfaction and employee turnover intentions.}

\medskip
\noindent\textbf{Perceived Job Alternative} \textsuperscript{*}

\begin{center}
\small
\setlength{\tabcolsep}{2.5pt}
\renewcommand{\arraystretch}{1.15}

\begin{tabular}{|>{\raggedright\arraybackslash}p{0.58\textwidth}|*{5}{>{\centering\arraybackslash}p{0.08\textwidth}|}}
\hline
 & \textbf{Strongly Disagree} & \textbf{Disagree} & \textbf{Neutral} & \textbf{Agree} & \textbf{Strongly Agree} \\
\hline
a) How easy would it be for you to find a job with another employer in this geographical area that is as good as the one you now have? & $\circ$ & $\circ$ & $\circ$ & $\circ$ & $\circ$ \\ \hline
b) How easy would it be for you to find a job with another employer in this geographical area that is better than the one you now have? & $\circ$ & $\circ$ & $\circ$ & $\circ$ & $\circ$ \\ \hline
\end{tabular}
\end{center}
\vspace{1\baselineskip}
\noindent\textbf{Please choose how perceived job alternatives mediates or moderates the relationship between job satisfaction and employee turnover intentions.}

\medskip
\noindent\textbf{Perceived Job Alternative} \textsuperscript{*}

\begin{center}
\small
\setlength{\tabcolsep}{2.5pt}
\renewcommand{\arraystretch}{1.15}

\begin{tabular}{|>{\raggedright\arraybackslash}p{0.58\textwidth}|*{5}{>{\centering\arraybackslash}p{0.08\textwidth}|}}
\hline
 & \textbf{Strongly Disagree} & \textbf{Disagree} & \textbf{Neutral} & \textbf{Agree} & \textbf{Strongly Agree} \\
\hline
a) How easy would it be for you to find a job with another employer in this geographical area that is as good as the one you now have? & $\circ$ & $\circ$ & $\circ$ & $\circ$ & $\circ$ \\ \hline
b) How easy would it be for you to find a job with another employer in this geographical area that is better than the one you now have? & $\circ$ & $\circ$ & $\circ$ & $\circ$ & $\circ$ \\ \hline
\end{tabular}
\end{center}
\vspace{1\baselineskip}
\noindent\textbf{Job Satisfaction} \textsuperscript{*}

\begin{center}
\small
\setlength{\tabcolsep}{2.5pt}
\renewcommand{\arraystretch}{1.15}

\begin{tabular}{|>{\raggedright\arraybackslash}p{0.52\textwidth}|*{5}{>{\centering\arraybackslash}p{0.09\textwidth}|}}
\hline
 & \textbf{Strongly Disagree} & \textbf{Disagree} & \textbf{Neutral} & \textbf{Agree} & \textbf{Strongly Agree} \\
\hline
In general, I like my job & $\circ$ & $\circ$ & $\circ$ & $\circ$ & $\circ$ \\ \hline
In general, I am satisfied with my job & $\circ$ & $\circ$ & $\circ$ & $\circ$ & $\circ$ \\ \hline
In general, I like working here & $\circ$ & $\circ$ & $\circ$ & $\circ$ & $\circ$ \\ \hline
\end{tabular}
\end{center}
\end{document}